\documentclass[a4paper]{article}
\usepackage[utf8]{inputenc}
\usepackage{amsmath}
\usepackage{amssymb}
\usepackage[T1]{fontenc}
\usepackage[margin=2.75cm]{geometry}
\usepackage{doi}
\usepackage[pdftex]{graphicx}
\usepackage{authblk}
\usepackage{graphicx}
\usepackage[normalem]{ulem}
\usepackage{multicol}
\usepackage{abstract}
\usepackage{sectsty}

\DeclareMathOperator{\Tr}{Tr}

\newcommand{\neswarrow}{\mathrel{\text{$\nearrow$\llap{$\swarrow$}}}}
\newcommand{\nwsearrow}{\mathrel{\text{$\nwarrow$\llap{$\searrow$}}}}

\title{Quantum information}
\author{Ryszard Horodecki\footnote{e-mail: ryszard.horodecki@ug.edu.pl}}
\date{}

\sectionfont{\fontsize{10}{12}\selectfont}
\subsectionfont{\fontsize{9}{11}\selectfont}

\affil{
  \small\emph{
    International Centre for Theory of Quantum Technologies, University of Gdańsk, Wita Stwosza 63, 80-308 Gdańsk, Poland
    and Institute of Theoretical Physics and Astrophysics, National
    Quantum Information Centre Faculty of Mathematics, Physics and
    Informatics, University of Gdańsk, Wita Stwosza 57,80-308 Gdańsk,
    Poland} }

\begin{document}

\maketitle

\begin{minipage}{8cm}
\vspace{-0.7cm}
\begin{quotation}
    \small\noindent
    Dedicated to memory of Roman Stanisław \mbox{Ingarden}
    on his centennial birthday
    \\[0.5em]
      ``\emph{…the quantum information theory is not only
        scientifically interesting subject, but is a practical need}''
      \begin{flushright}
        R.S. Ingarden\hspace{0.5em}\mbox{}
      \end{flushright}
\end{quotation}
\end{minipage}

\begin{abstract}
This article reviews the extraordinary features of quantum information predicted
by the quantum formalism, which, combined with the development of modern quantum technologies, have opened new horizons
in quantum physics that can potentially affect various areas of our live, leading to new technologies such as quantum
cybersecurity, quantum communication, quantum metrology, and quantum computation.

\medskip\noindent
topics:
quantum cryptography, quantum entanglement, nonlocality, entanglement witness
\end{abstract}

\medskip

\begin{multicols}{2}

\section{Introduction}

The concept of quantum information was born on the border between
quantum mechanics and information theory science. The stunning
success of the former has led to think that the concept of information cannot be separated from the mathematical
structure of quantum formalism that imposes fundamental constraints on the form of physical laws.

Already in the 1930s, von Neumann defined entropy \cite{Neumann} for quantum states as an
analogue of the classical Boltzmann-Gibbs entropy, which later turned out to be the quantum counterpart of Shannon
entropy \cite{Shannon} -- the concept underlying of classical communication theory. At about the same time, Einstein
Podolsky and Rosen pointed out the unusual features of quantum formalism that seemed to lead to the conclusion that
quantum mechanics is incomplete \cite{epr}.
In 1970, two young physicists,
Park \cite{Park} from the Department of Physics at Washington State University and
Wiesner \cite{Wiesner} from Columbia University in New York, independently analyzed the
physical implications of quantum formalism. While the former discovered a fundamental
limitation on copying quantum information, the latter discovered the first application of
quantum information to unforgeable quantum money. Unfortunately, both discoveries
were ahead of their time and passed unnoticed.
Three years later Holevo proved \cite{holevo} that there
is a bound for our ability to access classical information from
quantum systems which confirmed earlier Gordon’s \cite{Gordon} and
Levitin’s \cite{Levitin} conjectures.  This strengthened the conviction that Shannon's
communication theory is incomplete, in a sense that it did not
consider the transmission of all physical information carriers such as
quantum particles.  A few years later, Ingarden, a Polish
mathematical-physicist, published a work entitled:
``\textbf{\emph{Quantum information theory}}'' in
which he proposed a quantum generalization of Shannon's theory
in terms of the generalized quantum mechanics of open systems \cite{Ingarden-qit} (see also \cite{IKO}).
However, it was only a series of seminal papers \cite{Bell,
  Aspect1982, BB84, Deutsch, Feynman, E91, BennettW, Deutsch_Jozsa,
  Teleportation, ent_swapping, Shor, Shor_alghor, S_qubit,
  Grover_97, Wieden-teleportation, Rzym-teleportation,
  Kimble-teleportation} that revealed specific features the quantum
code of nature pointing to the quantum origins of information.

There were various reasons for the
relatively late advent of the quantum information era crowned with the bilding of
Shannon's quantum theory (see \cite{Wilde}).  In particular, the unusual
success of
Shannon's theory led to the belief that the laws of physics could be derived from information processing as a purely
mathematical concept detached from physical information carries.
On the other hand the identification of peculiar features of quantum
information such as monogamy of entanglement \cite{Bennett_capacity, Kundu,
  Terhal_PToday} required advanced quantum technologies.
Additionally, the obstacle was the abstract, mathematical and
non-intuitive nature of the standard quantum formalism, which looked
like inscription, not all predictions of which were entirely clear
even for its fathers.

\section{Quantum inscription as a paradigm for quantum information}

Roughly speaking, quantum inscription is an instruction -- a set of prescriptions that
determine the way of probabilistic prediction of the results of future measurements in laboratories
\cite{Nielsen-Chuang, Breuer_OpenQS}.

Each physical system corresponds to complex vector space Hilbert $H$ equipped with the
linear scalar product $\langle.|.\rangle$ such that the space is complete with respect to the norm
\begin{equation}
  \|\psi\| = \sqrt{\langle\psi|\psi\rangle}.
\end{equation}
The space $H$ of system $S$ compound of $n$ subsystems $S_1, S_2, \ldots, S_n$ is a tensor product
$H = H_1 \otimes H_2, \ldots \otimes H_n$ of the Hilbert space of subsystems. The
subsystems can represent distinguishable particles, various complex objects, e.g. atoms, molecules, or different
degrees of freedom of the same object, e.g. photon polarization and propagation modes.

The central object is the wave function (state vector) $|\psi\rangle$ with
the unit norm $\|\psi\|=1$, which is an element of a Hilbert space. It
contains all probabilistic information about the system and satisfies the Schrödinger equation:
$i\hslash{\partial|\psi\rangle \over \partial t} = H|\psi\rangle$, where $H$ is linear
self-adjoint operator called Hamiltonian.
The symbol $\varrho$ denotes the state of the system about which we only have partial information.
It can be described by a Hermitian positive semidefinite operator with unit trace:
$\varrho=\varrho^\dagger$, $\varrho\geq 0$, $\Tr(\varrho)=1$
where trace $\Tr(\varrho) = \sum_k \langle\phi_k|\varrho|\phi_k\rangle$ and sum runs
over diagonal elements in arbitrary orthonormal basis $\{\phi_k\}$. The symbol $U$ stands for
unitary operations that transform states, and in the case of pure states, they keep the
scalar product preserved.

Observable quantities correspond to Hermitian linear operators $O$ acting on the state space
$H$.
In contrast to classical observables the quantum ones can be noncommutative:
$[O_1, O_2] = O_1O_2 - O_2O_1 \neq 0$. The
most familiar example is $[Q, P] = i$ where $Q$ and $P$ are the
position and momentum
operators. The structures and mutual interrelations of noncommutative observables bring deep questions
concerning the properties of the quantum systems related to the fundamental principles: uncertainty and
complementarity. The first one limits the precision of the statistics of the results of two complementary observables,
such as position and momentum \cite{Heisenberg}.   The complementarity
principle says that two quantum observables cannot be measured
simultaneously, and thus provide ``independent'' information about physical systems \cite{Bohr1}.

Contrary to classical theories, quantum measurement is active. It creates properties, does
it randomly, and can change state if the latter is not specially tailored for a given measurement. The measurement does not
always provide information about state but it can be part of a quantum operation. Any state $\varrho$ defines the
probability distribution as the mapping assigning to each measurement result $i$ the probability $p_i$ of that
measurement result (the Born rule):
\begin{equation}
  p_i = \Tr [\Pi_i \varrho]
\end{equation}
where $\{\Pi_i\}$, $\sum_i \Pi_i = \boldsymbol{I}$ are elements of a positive operator-value measure (POVM) and $\boldsymbol{I}$ is
unit operator. In particular, if $\Pi_i$ is projector operator then the generalized measurement
correspond to the von Neumann measurement, which completely determines the post-measurement state.
After the measurement with the outcome $i$,
the system goes to the post-measurement state
\begin{equation}
  \varrho’_i = p_i^{-1} \Lambda_i (\varrho)
\end{equation}
where $\Lambda_i (\varrho) = \Pi_i \varrho \Pi_i$ is particular positive
superoperator which clearly maps positive operators to positive operators and normalization of
$\varrho'_i$  requires the condition to be met
$\Tr [\varrho \Pi_i] = \Tr [\Lambda_i (\varrho)]$ where $\Lambda_i (\varrho) = p_i \varrho_i'$.
The most general physically implementable map is
a completely positive map $\Lambda$ which satisfies condition:
$\Lambda \otimes \boldsymbol{I}_n \in B(H_1 \otimes C^n, H_2 \otimes
C^n)$, where $B$ is space of positive maps between the Hilbert spaces
$H_1 \otimes C^n$ and $H_2 \otimes C^n$, $\boldsymbol{I}_n$ is unit
operator on $n$ dimensional Hilbert space $C^n$.  If in addition
$\Lambda$ is trace-preserving it
determines quantum channel which play a central role in the processing of quantum
information \cite{Wilde}.
Any completely positive map on a system $S$ in a given state $\varrho$ can be
realised via unitary interaction of $S$ with some other system (ancilla)
in a pure state followed by von Neumann measurement and final partial
trace.  This fact comes from so called Stinespring dilation theorem
\cite{Nielsen-Chuang}.

The crucial difference between the
quantum description of physical reality and the classical one is
the principle of superposition: if $|\Psi_1\rangle$, $|\Psi_2\rangle$ are system states then their superposition;
\begin{equation}
  |\Psi\rangle =  a|\Psi_1\rangle + b|\Psi_2\rangle
\end{equation}
is also in good state, provided that $a$ and $b$ are chosen so that
$|\Psi\rangle$ is normalized.

The prediction power of quantum inscription is astonishing:
``\textbf{All our experience so far using quantum theory seems to say:
What is predicted by quantum formalism must come to the laboratory}''
\cite{RMP2009}.
In the early 1970s, it seemed that all possible predictions of quantum inscription had already
been recognized. The papers of Einstein, Podolsky and Rosen \cite{epr} and Schrödinger \cite{Schrodinger1} were initially
treated rather as a mathematical artefact detached from its physical implications. Ironically, it was them who drew
attention to the extraordinary implications of quantum inscription, which revealed the existence at a fundamental level
of a subtle order governed by quantum information. In the classical world, quantum information is
``unspeakable''. It cannot be written with discrete symbols, e.g. on a tape of a Turing machine.
So far, there is no commonly accepted definition of quantum information.

For our purposes, it is convenient to adopt the following
interpretation: \textbf{\emph{Quantum information is what is carried by quantum particles
    and the wave function $\boldsymbol{\psi}$ is its mathematical image}} \cite{IBM_2004}.

\emph{Quantum information} (QI) can be processed (manipulated) \cite{Nielsen-Chuang,Raimond2001},
using combinations of unitary operations and measurements. QI is the source of quantum
resources \cite{Gour_RMP} such as entanglement
\cite{RMP2009,AmicoRMP,TothPhysRep}, steering \cite{WJD_stiringEPR}, quantum
correlation beyond entanglement \cite{Modi_RMP}, quantum coherence
\cite{S_coherence_RMP}, asymmetry \cite{Gold_Asymmetry}.
It allows to perform nonclassical
tasks such as quantum cryptography \cite{BB84, E91, Robust_number, Gisin_RMP},
teleportation \cite{Teleportation, Wieden-teleportation, Rzym-teleportation, Kimble-teleportation},
quantum computing \cite{Feynman, Deutsch, Deutsch_Jozsa},
not feasible with classical resources.  QI is resource for quantum metrology
\cite{RMP_metr}, computational complexity \cite{Buhrman2010,Zuk_compl,Bour_complex}.

However, this subtle resource has a very unpleasant feature.  As one knows, non-diagonal
elements of the density matrix $\varrho$ called coherence
in the state $\varrho$, provide information about quantum interference.
Unfortunately, as a result of the system's
interaction with the environment, the process of decoherence \cite{Zurek_decoher_RMP} occurs, which causes
disappearance of non-diagonal elements of density matrix of the state. Reversing the degradation of quantum information still
remains a great challenge for effective processing of quantum information.

\section{Quantum bit -- the unit of quantum information}

The concept of qubit appeared for the first time in the context of the theory of quantum
information transmission \cite{S_qubit} as a two-level system, the state of which can be written as a superposition of two
base states $|0\rangle$ and $|1\rangle$
\begin{equation}
  |\Psi\rangle = a|0\rangle + b|1\rangle
\end{equation}
where $a$ and $b$ are complex numbers, $|\Psi\rangle \in \boldsymbol{C}^2$ (two-dimensional Hilbert space).

Contrary to the classical bit, the qubit represents a continuum of possible states defined by
its wave function, which can be visualized by the two-dimensional Bloch sphere with two real parameters
$\theta$ and $\varphi$ where $a = \cos(\theta/2)$, $b = \sin(\theta/2 \exp(i\varphi))$ 
where $0 \leq \theta \leq \pi$, $0 \leq \varphi \leq 2\pi$.
For illustration, consider a photon as a
paradigmatic example of a qubit. It requires a Hilbert space $H$
which is a tensor product $H = H_{prop} \otimes H_{pol}$, where
$H_{prop}$ represents the photon
propagation modes while $H_{pol} = \boldsymbol{C}^2$
describes the photon polarization modes. If one disregards the propagation modes, the photon can be treated as a
photonic qubit in polarization degree of freedom.

Consider now a photon in the superposition of the base states
$|0\rangle \equiv |\!\!\updownarrow\rangle$,
$|1\rangle \equiv |\!\!\leftrightarrow\rangle$
corresponding to vertical and horizontal polarization
$|\Psi\rangle = \sin \Theta \, |0\rangle + \cos\Theta \, |1\rangle$.  If we direct it to a vertical polarizer,
it will change to one of the states $|0\rangle$ or $|1\rangle$ with probabilities
$p_0 = Tr[\Pi_o \varrho] = \sin^2\Theta$, $p_1 =  Tr[\Pi_1 \varrho] = \cos^2\Theta$ respectively,
where $\Pi_0  = |0\rangle,\langle0|$, $\Pi_1 = |1\rangle,\langle1|$ are projectors
and density matrix of the state $|\Psi\rangle$ is given by
\begin{equation}
  \varrho = |\Psi\rangle\langle\Psi| =
  \begin{bmatrix}
    \sin^2 \theta & \sin\theta\cos\theta \\
    \sin\theta\cos\theta & \cos^2\theta
  \end{bmatrix}
\end{equation}
where the diagonal elements are interpreted as the probabilities of the basis state, while
the off-diagonal elements represent the coherence of the basis states.

\begin{figure*}
  \centering
  \includegraphics[width=8cm]{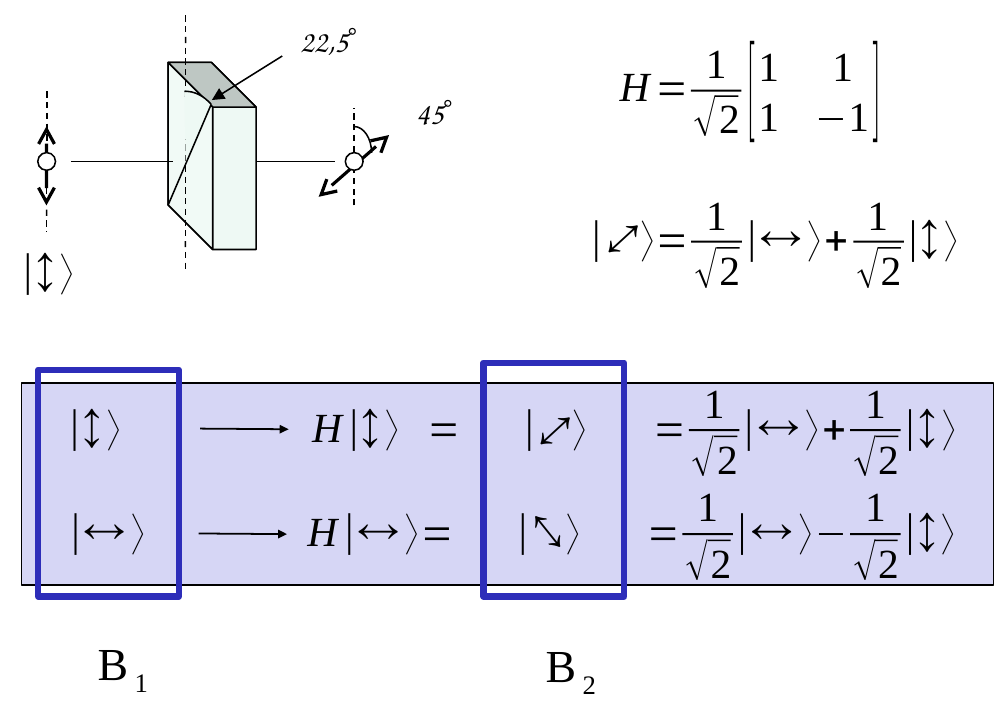}
  \caption{The photonic realization of the Hadamard $H$ gate.
  $\boldsymbol{B}_1$ and $\boldsymbol{B}_2$ denote the computation base and the Hadamard base.}
  \label{fig:hadamard}
\end{figure*}

If we now place a specially cut
birefringent crystal with the optical axis at an angle of 22.5 degrees on the path of a vertically polarized photon,
the photon will be in a state of linear superposition (Fig. \ref{fig:hadamard}).
This is nothing but the photonic realization of the Hadamard $H$
gate. It has no classical counterpart and plays a fundamental role in quantum information processing including quantum
computing. Note that arbitrary photonic wave plate operations for
photonic polarization qubits realizing Hadamard, Pauli-$X$, and rotation gates were
implemented on the chip \cite{HadamardGate}.

In Fig. \ref{fig:hadamard}, $\boldsymbol{B}_1$ and $\boldsymbol{B}_2$
denote the computation base and the Hadamard base, respectively, which are
mutually unbiased, i.e. they are mutually exclusive. Perfect information about the polarization along the selected axis
implies that there is no information about the polarization along the axis rotated by $45^{\circ}$. This is a purely quantum
mechanical effect resulting from the fact that the vectors $|0\rangle$, $|1\rangle$ and $|+\rangle \equiv |\!\!\neswarrow\rangle$, $|-\rangle \equiv |\!\!\nwsearrow\rangle$
are the eigenstates of the Pauli operators $\sigma_z$ and $\sigma_x$, respectively, which do not commute, ie.
$[\sigma_z, \sigma_x] = \sigma_z\sigma_x - \sigma_x\sigma_z \neq 0$.

There have been many proposals for the physical
realization of a qubit on quantum dots \cite{quantum_dots} electron spins \cite{electron_spin_qubit},
semiconductor spin \cite{Press_semicon_spin}, superconducting
charge qubits based on Josephson junction \cite{Supercond_qubits, Qubits_Joseph._junct}.
Remarkably it has been demonstrated, that linear optics is sufficient for efficient quantum
information processing with photonic qubits in two optical modes (such as horizontal or
vertical polarization) \cite{KLM_comp, Kok_RMP_comp}.  There has recently taken place a quite progress in
parallelized quantum information processing which includes tailored quantum memories
to simultaneously handle multiple photons \cite{Parniak_Nature}.

\section{Fundamental limitations on quantum information processing}

Already in 1961 Wigner pointed out that the existence of self-reproduction in the
quantum world is unlikely \cite{Wigner}.
In 1970. Park \cite{Park}, and later Wooters and Żurek \cite{WZ-no-clon} and
Dieks \cite{Dieks_no-clon} proved that it is impossible to build a quantum machine that can perfectly
copy arbitrary unknown quantum state $\Psi$:
\begin{eqnarray}
  |\Psi\rangle |0\rangle |M\rangle &{\not\to}& |\Psi\rangle |\Psi\rangle |M_\psi\rangle
\end{eqnarray}
where $|0\rangle$ means a blank state, while $|M\rangle$, $|M_\psi\rangle$ are machine state before and after
cloning respectively.
The process realized by such a machine would have to be nonunitary and non-linear, which is
forbidden by the linearity of quantum formalism. Thus copying destroys the state and it cannot be reconstructed from a
single copy. Hence the quantum signals cannot be noiselessly amplified.
Later
the limitation for the unperfect cloning in terms of the so called fidelity function
$f(\varrho_{out}) = \langle\Psi| \varrho_{out}|\Psi\rangle$ measuring
similarity of the state of either of the two outcome registers has been provided within the framework of imperfect quantum
cloning machines \cite{no_perfect_clon, Fan2014}. There is dual the non-deleting  theorem,
which states that, in general, given two copies of some arbitrary quantum \\
state, it is impossible to delete one of the copies
\cite{KumarPati2000}.
In the above mentioned paper Holevo \cite{holevo} proved an fundamental
theorem that sets an upper limit to the amount of information available about a quantum state. It implies
that with the help of one qubit is impossible to send more than one bit of classical information.

Quite unexpectedly, it turned out that there is also a restriction on the possibility of
generating of quantum superposition. Namely, it has been independently
shown \cite{AlvarezRodriguez2015, Oszmaniec2016} that there is no
universal probabilistic
quantum protocol generating superposition of the two unknown states. Interestingly, a probabilistic protocol generating
a superposition of two unknown states having a fixed overlap with a known pure reference state has been proposed
\cite{Oszmaniec2016}.
This protocol has been carried out experimentally in a three-quadrant NMR system as well as on unknown
photonic quantum states \cite{Li2017, Hu2016}.

\section{Quantum cryptography based on no-cloning}

Parallel to Park's paper on non-cloning, Wiesner, based on principle
of uncertainty introduced the concept of conjugate coding to make up
quantum money \cite{Park}. This idea paved the way for the quantum information
encryption Bennett’s and Brassard’s protocol (BB84) \cite{BB84}. It has the
following main three steps:

\begin{enumerate}
\item Alice sends randomly polarized photons through the quantum channel in the
selected computing bases $\{B_1\}$ $|0\rangle$, $|1\rangle$ and Hadamard $\{B_2\}$ $|+\rangle$, $|-\rangle$; saves bases and bits.
\item Bob measures photons in randomly selected bases $B_1$ and $B_2$, registers bases and bits.
\item Via the classic public (authenticated) channel, Alice and Bob
  transmit their choices bases. When their bases match, they retain the
  appropriate bits.
\end{enumerate}

Thus, they receive a raw key that requires further processing. To check for eavesdropping,
they calculate the quantum bit error of a randomly selected data
subset that they reveal each other via the public channel and check if
the error
(percentage of mismatched bits) is below a certain
threshold value. Using classic post-processing protocols such as error correction and privacy amplification, they
generate the final secure key.

Since 1992, when Bennett and Brassard and colleagues demonstrated the first 32cm quantum
distribution of the key in free space \cite{Prototyp_gen},
there has been tremendous progress in the development of quantum cryptography
in free space and in fiber. There is a continuous improvement of cryptographic keys over long distances \cite{QKD_G-Lozanna}
as well as an
increase in key generation speed using single photon detectors \cite{Wengerowsky}. Quantum key
distribution (QKD) networks were established in the US, Austria, Switzerland, China and Japan. and the European SECOQC
network \cite{SECOQC}. Due to exponential signal attenuation and decoherence, the effective distribution range of the quantum key
of terrestrial networks is limited to 300 km \cite{Korzh300km}. In cosmic space, both of these factors are many times weaker. In 2016, the first
satellite distribution of the BB84 protocol was performed using a one-time key cipher via the Micius satellite at
intercontinental distances, thanks to which the photos of Schrödinger and the philosopher Micius were safely
transferred between Vienna and Beijing \cite{Micius}.

Despite the enormous advances in quantum cryptography, there are still some problems
related to the fact that practical implementations of quantum key decomposition use realistic photonic qubits and
imperfect single photon detectors. This creates gaps between QKD theory and practice enabling quantum hacking, e.g. The
Bright illumination Attack, Photon number splitting \cite{Tal_gaps}. Therefore, QKD implementations are still in the testing
phase and these gaps are identified. Stronger versions of BB84 were developed, such as the BB84 decoy state and
protocols resistant to photon number breaking attacks \cite{Robust_number}.
As a result, QKD protocols become more and more secure.

\section{Quantum entanglement -- the most non-classical feature of quantum information}

As we have seen, already at the level of simple systems, the properties of quantum information
differ substantially from those of classical information that can be amplified and copied. Much earlier, in the 30s,
EPR and Schrödinger revealed a peculiar feature of quantum information in complex quantum systems rooted in the
principle of superposition called entanglement. According to the quantum inscription, the state space
$H_S$ of the quantum system $S$ compound from distinguishable subsystems
$S_1, S_2, \ldots S_n$ is given by $H_{S1} \otimes H_{S2}, \ldots \otimes H_{Sn}$
which is the tensor product of the Hilbert space of the subsystems.

We say that a pure state is entangled if it cannot be written as the product of the states of the
individual subsystems
\begin{equation}
  |\Psi\rangle_{12\ldots n} \neq |\phi\rangle_1 \otimes |\psi\rangle_2\ldots \otimes |\chi\rangle_n
\end{equation}

In general a mixed state $\varrho$ of $n$ systems is entangled if it
cannot be written as a convex combination of product states
\begin{equation}
  \varrho \neq \varrho_{sep} = \sum_i p_i \varrho_1^i \otimes \cdots \otimes \varrho_n^i
\end{equation}

In particular, for any two-part pure entangled state $|\psi\rangle_{12} \in H_1 \otimes H_2$
there exist orthonormal Schmidt bases
$\{\phi_i\rangle$, $\{\chi_i\rangle$ in $H_1$, $H_2$ respectively such that:
\begin{equation}
  |\Psi\rangle_{12} = \sum_i^d c_i |\phi_i\rangle \otimes |\chi_i\rangle
\end{equation}
where the summation takes place on the smaller dimensions of the two systems $d = \min(d_1,d_2)$.
In particular, the two-part maximally entangled state in the space
$H_1 \otimes H_2$ with the dimension $d^2$ is defined as:
\begin{equation}
  |\Psi_{\max}\rangle = \frac1{\sqrt{d}} \sum_i^d |\phi_i\rangle \otimes |\chi_i\rangle
\end{equation}

In particular there is a two qubit entangled state:
\begin{equation}
  |\Phi^+\rangle = \frac1{\sqrt2} ( |0\rangle_1 |0\rangle_2 + |1\rangle_1 |1\rangle_2)
\end{equation}
where $\{|0\rangle$, $|1\rangle\}$ is the computational basis for
qubits. Using von Neumann entropy as a measure of entanglement for pure states, it is easy to check that the above state
above contains one ebit of entanglement, i.e. the maximum amount of entanglement that a system with dimension $d=2^2$
allows. In general, for a system consisting of $n$ pairs of entangled qubits and a Hilbert space dimension, $d=2^n$
contains $n$ ebits of entanglement. Most of the pure state vectors in a pure state two-part Hilbert space are not maximally entangled.

For systems divided into more than two parts, the Schmidt distribution in general does not
exist. However, many of the important states in quantum information processing take the form of a multi-part Schmidt
distribution. Among them, three-particle W and GHZ states: $|\Psi\rangle_{GHZ}=(|000\rangle + |111\rangle)/\sqrt2$ \cite{GHZ},
$|\Psi\rangle_W=(|001\rangle + |010\rangle + |100\rangle)/\sqrt3$ \cite{W}, which represent two different types of
entanglement that cannot be transformed into each other through local operations and classical communications (LOCC).
Interestingly experimental W-to-GHZ state conversion were recently demonstrated \cite{Zheng, Haase}.

Let us emphasize that the above mathematical description of quantum entanglement between the
various degrees of freedom of complex systems is adequate in a scenario where each subsystem (e.g. qubit) can be
individually addressed / manipulated. In situation when one consider indistinguishable systems in connection with
symmetrisation postulate the complete characterization of entanglement is still challenge. Many different approaches
have been proposed with different entanglement definitions. Recently, Benatti et al. \cite{Benatti_ent.indist}
made an extensive
comparative analysis of different approaches to the definition of entanglement of quantum systems composed of
indistinguishable particles based on natural physical requirements.

There are many ways to generate quantum entanglement.
Entangled states are most often generated in the spontaneous
parametric down-conversion and spontaneous four-wave mixing \cite{Kwiat_95, Erhard_2020, Akopian_PRL2006}.
It is intriguing that it
is possible to entangle together particles from two independent
sources that did not interact with each other in the past
\cite{Yurke_Stoler,ent_swapping}. Another peculiar behaviour of
entanglement called a sudden entanglement death was described in a
dynamic scenario. Namely, when two entangled qubits interact with
natural reservoirs, the entanglement can disappear in a finite time
while the coherence disappears asymptotically \cite{Sadden_Death, Eberly1,
Eberly2}. The source of this phenomenon is due to the fact that in
finite-dimensional systems the set of separable (non-entangled) states has a
finite volume \cite{Z_Pawel_Maciek}.
This important result was in
particular discussed in the context of quantum computing on NMR which
operates on highly mixed, separable states \cite{Kus_Z}.

The discovery of Einstein, Podolsky, and Rosen that entangled states could show
``ghostly'' correlations independent of distance, until the appearance of John Bell's famous
work, was not given much interest. On the one hand, they were considered more philosophical than physical, on the other
hand, it was believed that such correlations could be simulated classically.

\section{Photons entangled in polarization}

\begin{figure*}
  \centering
  \includegraphics[width=10cm]{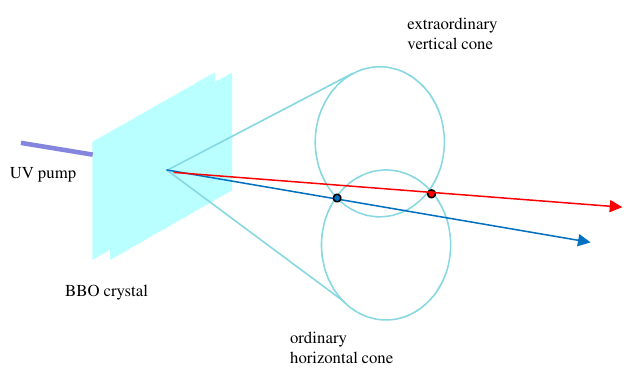}
  \caption{Generation of photons entangled in polarization using
    Type-II conversion in the Bell state $|\Psi^+\rangle_{AB}$}
  \label{fig:2}
\end{figure*}

To illustrate this phenomenon, consider the probabilistic generation of photons entangled in
polarization degrees of freedom using Type-II down-conversion \cite{Kwiat1995}.
In this process, a high-energy photon in an optical nonlinear medium (BBO crystal) is converted into two
lower-energy photons that are emitted along the surface of two anti-correlated intersecting cones with vertical and
horizontal polarization (Fig. \ref{fig:2}). In particular, the photons
emitted along the intersections cannot be assigned a specific
polarization because we do not know which cone they come from. We write it down as a quantum alternative
\begin{equation}
  \label{eq:12}
  |\Psi^+\rangle_{AB} = {1 \over \sqrt2} ( |0\rangle_A |1\rangle_B + |1\rangle_A |0\rangle_B),
\end{equation}
where $|0\rangle$, $|1\rangle$ correspond to vertical, and horizontal polarisation respectively.
Here $|\Psi^+\rangle$ is one of four canonical Bell-states (Bell basis) \cite{Mann_Bell}:
$|\Psi^\pm\rangle = {1 \over \sqrt2} ( |0\rangle_A |1\rangle_B \pm |1\rangle_A |0\rangle_B)$,
$|\Phi^\pm\rangle = {1 \over \sqrt2} ( |0\rangle_A |0\rangle_B \pm |1\rangle_A |1\rangle_B)$,
Now, if we direct the photons from the entangled pair (EPR pair) in the polarization to the distant Alice's
and Bob's laboratories, respectively, who independently measure the polarization of the
same type, it turns out that they get anti-correlations $0-1$ or $1-0$.  What is striking is the fact
that individual photons do not carry any bit because their polarization is completely random \cite{E91}, so
local measurement results turn out to be completely random too. EPR reasoned as follows: If it is possible to
``remotely'' predict some property of a particle without interacting with it, then this
property must have existed before, ie. before the measurement. They called it the ``reality elements'', and from there they
concluded that quantum inscription offered an incomplete description of physical reality.

\section{Nonlocality of quantum correlations. Bell tests}

It was a serious objection that no one, including Bohr himself, was able to convincingly
refute. The Gordian knot was cut by John Bell \cite{Bell}. Namely, he
formalized the concepts of reality elements by introducing a
model of local hidden variables based on the following assumptions: i) the measurement results are determined by the
properties of the particle carried before and independent of the measurement, ii) the results determined in one place
are independent of any actions in the space-like separation, iii) the settings of local apparatus do not depend on
hidden variables that determine the results of local measurements. These assumptions, as Bell showed, impose constraints on
correlations called Bell's inequalities. The key point is that they can be verified in the laboratory regardless of any
theory.

Let us briefly illustrate the Bell inequalities by exemplifying the correlation of polarized
entangled photons that were sent to the distant Alice and Bob laboratories along the z axis. The partners measure
dichotomous observables ie polarizations that have only two values $+1$ or $-1$. Each partner measures two
such observables. Alice chooses the settings of detectors $a$, $a'$, Bob $b$, $b'$, which are unit vectors showing
different angles in the x-y plane along which they can orient polarizing filters. For each pair of settings,
correlation functions can be constructed: $\langle a, b\rangle$, $\langle a, b '\rangle$,
$\langle a', b\rangle$  $\langle a ', b'\rangle$, where $\langle \rangle$  means the average of
the product of outputs. On this basis, it is possible to build a new Bell observable $B = a, b + a, b' + a', b + a',
b'$. Now, if we accept the assumptions of local realism, in particular that each photon had a
certain polarization value ($+1$ or $-1$) before the measurement, it is easy to check that the absolute value of Bell observable
cannot exceed 2. Hence, we obtain Bell-CHSH inequality \cite{CHSH}:
\begin{equation}
  \label{eq:13}
  |\langle ab+a'b + ab' - a'b'\rangle_{kl}| \equiv |\langle B \rangle| \leq 2
\end{equation}

Quantum mechanics predicts that the mean value of the $B$ observable satisfies the inequality
$|\langle B_{QM}\rangle| \leq 2\sqrt2$ which means that it breaks Bell-CHSH inequality, where $2\sqrt2$ is the
so-called Tsirelson's bound \cite{Cirelson1980}.

The verification of Bell's inequality based on the assumptions of local realism proved to be a
great challenge for experimentalists, as it required the closure of three loopholes:
i) Locality demands that no signal traveling at the speed of light can inform the distant
detector of its settings or the result of a measurement on the local detector before Alice
and Bob complete the measurements;
ii) Fair-sampling (or detector efficiency) demands that the sample of
entangled pairs be a faithful representation of the entire ensemble being broadcast; iii)
Freedom of choice requires that the hypothetical local variable should not influence the local
choices of measurement setups on the part of Alice and Bob.

The first ground breaking experiment that convincingly demonstrated breaking the Bell CHSH
inequality and good agreement with the predictions of quantum
mechanics was performed by Aspect et al. \cite{Aspect1982}.
In their experiment,
entangled photon pairs were emitted by the process of atomic calcium cascades. For the first time, the authors used
acousto-optical switches, which pseudo-randomly changed the orientation of the analyzers in a short time compared to
the photon transit time and detection. They achieved more than 95\% of the detection efficiency.

Only in 2015, a series of Bell tests based on quantum random number generators was performed,
which closed both locality and fair-sampling loophole in the same experiments \cite{LoopholeKaiser}.
Recently, two cosmic Bell tests with photons entangled in polarization were performed, in
which measurement settings were determined by real-time photon wavelength measurements from high redshift quasars,
light emitted billions of years ago; Thus, the authors closed two loopholes at once: locality and freedom of choice
\cite{Loophole_Zeiling, LoopholeGisin}. However, these experiments failed to close the fair-sampling loophole.
Quite recently Pan et al. \cite{LoopholePan}  performed an impressive local realism test that closes both
locality and fair-sampling loophole and rules
out common cause 11.5 years before the experiment, which largely closes the freedom of choice loophole.

The interpretation of violating Bell's
inequality is still the subject of the discussions \cite{Zukowski_Brukner,Cavalcanti}. The
Bell tests show that the quantum correlations cannot be explained using any theoretical
model based solely on local variables.
This particular feature of quantum information, which has become known
as quantum nonlocality (Bell nonlocality), provides the resource for device-independent quantum
key distribution \cite{QKD, Brunner_RMP, ReDIQIP} (see however \cite{Acin2021}).

\section{Weaker forms of breaking realism}

While I'm not going to do a detailed review of the vast field of difference in Bell's inequality, let me mention two
important related concepts.  First, it should be mentioned that violation of local realism by composed quantum
systems has it’s a weaker quantum analog called quantum contextuality, observed with help of random
measurements of specially designed sets of quantum measurements pioneered by \cite{Kochen_Spec} which has
many further developments (see \cite{Cabello, Cabello2008, Exp_conBoure}) can be mathematically quantified
\cite{QuantContext}.  Quite remarkably it have the so called state variant fully analogous to Bell inequalities
\cite{StateDepend} as well as state-independent one which is valid for any state, and basically reports the
nonclassicality of the sent of measurement involved \cite{Mermin}.

The fundamental difference is that roughly speaking quantum contextuality can contradict classical realism only
under assumption of some bound on dimension of Hilbert space, while violation of Bell inequalities via quantum
states is the phenomenon that is independent on that assumption in general. This is why violation of the
inequalities in many cases leads to the powerful concept of quantum \emph{self-testing} \cite{Self}. In the case of the
inequality \eqref{eq:13} self-testing means that independently of complexity of local systems (for instance one may
assume that each of the observables in \eqref{eq:13} may concern not polarisation but some other or even all of the
photon internal degrees of freedom) the saturation of the quantum bound $2\sqrt2$ guarantees that up to local
isometries and local partial traces the state is in the \emph{unique}
qubit form \eqref{eq:12}. This is an essence of the device independent
variant of the Ekert's entanglement-based encryption protocol (E91)
\cite{E91} (see Sec. \ref{sec:end-based-crypt}). Quantum self-testing is a cornerstone of
device independent quantum
cryptography which is based on the idea that only the output statistics of the devices are enough to guarantee
cryptographic security without need of knowing the physical structure of the devices (for example see
\cite{RandomNumbBell}. Finally there is a weaker variant of Bell inequalities on composite systems that is still much
stronger than contextuality. This is based on the so called quantum steering \cite{WJD_stiringEPR} in which we
assume that for one of the particles the dimension of the Hilbert space is known (much like in contextuality tests)
while in the other is not. This leads to the so called semi-device independent quantum cryptography (see
\cite{SemiDeviceInd} and reference therein), \cite{Steering_Gda}.

\section{Nonlocality and the principle of informational causality}

The discovery of quantum nonlocality shook our perception of the foundation of quantum
physics. Hence the natural question arose:
Is there a nonlocality stronger than that predicted by quantum formalism? Is this the only
description that allows for nonlocal phenomena consistent with special relativity?
In the 1994 paper, Popescu and Rohrlich (PR) \cite{PR, Popescu_NP} took nonlocality as the basic axiom and
have proposed a model independent approach, consistent with special relativity,
based on the conception of input-output black-box devices.  In the approach
the experiments of Alice and Bob are space-like separated and each experiment is treated as
a black-box. Then all the physical information obtained in the experiment is encapsulated in the joint probability
$P(a,b|x,y)$ that Alice obtains $a$ and Bob $b$ when Alice inputs $x$ and Bob inputs $y$ respectively.
In the simplest case where
$x$, $y$, $a$, $b$ have only two possible values, they must satisfy the constraints: $a \oplus b = xy$
where $\oplus$ denotes addition
modulo 2. It is not difficult to verify that PR nonlocality leads to algebraic breaking of CHSH inequality equal to 4
which drastically breaks Tsirelson's limit $2\sqrt2$.
Does nature allow information to be processed using such
super-quantum correlations? Remarkably the physical principle of
information causality was proposed \cite{Inform_caus}, which excludes
such possibility. The information causality principle can be formulated briefly as:
\textbf{\emph{The message cannot contain access to more information than the amount
contained in it.}}
Contrary to its laconic form, this principle has strong implications:

\begin{itemize}
\item It strictly determines the maximum value of quantum correlations $\leq 2\sqrt2$
\item it is fulfilled by both classical theories and quantum mechanics
\item it excludes the physicality of the super strong Popescu-Rohrlich correlations
\end{itemize}
It is significant that although the properties of quantum and classical information are
basically different, they both follow the principle of informational causality.
It should be noted here that nonlocal PR boxes although
nonphysical provide a conceptual tool in the modeling of nonlocality
in the quantum physics and beyond \cite{Piani_boxes, Karol_PR, Gisin_nonlocality}.
It is remarkable that the PR correlations are under some
circumstances much more powerful resource than quantum entanglement as they lead to trivialising
quantum communication complexity \cite{Dam_thesis, LimitNonlocal}.  However they are weaker in another sense
since in their language there is no room for nontrivial dynamics and continuous chance of settings of the
measurements.

Finally it is worth noting that in the case of three parties the
concept of relativistic causality that goes beyond the no-signaling
paradigm is possible when space-time variables are explicitly
involved \cite{Grunhaus1996,HorRam_NatCom}.
Quite recently the general axiomatic approach to causality of the
evolution of the spatial statistic detection has been initiated
\cite{Eckstein2020, Miller2021}.

\section{Entanglement-based cryptography}
\label{sec:end-based-crypt}

As mentioned above, quantum correlations, apart from nonlocality, have another feature -- they
are random. It was intriguing that this randomness ensures the peaceful coexistence of quantum inscription predictions
and special relativity, as partners cannot use the correlation to the instant telegraph. This specific
``telegraphic no-go'' has not yet had clear theoretical foundations, although recently an
attempt to explain this phenomenon has been made \cite{Dragan_Ekert}.

As we saw, singlet-state photon pairs entangled generate anti-correlated random numbers at
distant locations.  Ekert first noticed that the randomness of these correlations could be used to generate a
secure cryptographic key and proposed the protocol E91 \cite{E91} based on the
entangled spin$\frac12$ particles in singled state and Bell’s theorem and proposed implementation
using nonlocal correlations between maximally entangled photon-pairs.
Soon after, the Bennett, Brassard and Mermin proposed a simplified
protocol based on entanglement without Bell's theorem, and showed that
it is equivalent to BB84.
The security of E91 is due to the fundamental property called monogamy
of entanglement which express the fact that entanglement represents
correlations that cannot be shared by third parties \cite{Bennett_capacity, Kundu,
  Terhal_PToday}. This peculiar entanglement trait not only provides
the security of entanglement-based cryptography, but sheds new light
on physical phenomena in many correlated systems
\cite{Winter_mon}.

Experiment implementations of the E91 protocol have been made at ground stations \cite{Peloso2009,Fujiwara2014}.
Recently, both production and analysis of entangled states have been tested with the SpooQy
satellite, which is a step towards the realization of a cryptographic key generator based on
entanglement in cosmic space \cite{SpooQy}. Quite recently, the
quantum key distribution has been analyzed with a small block length,
which
is crucial in entanglement-based quantum communication \cite{EkertChin}.
It should be emphasized that the original E91 protocol was prophetic as it suggested device-
independent cryptography \cite{Barrett, Brunner_RMP}, based on Bell inequality breaking,
which ensures that the data produced by quantum devices has a certain degree of secrecy, no
matter how exactly the data was generated.

\section{Canonical effects based on quantum entanglement}

Ekert's work was important for another reason, namely, it was the first to show that ``ghostly''
EPR correlations can be harnessed into something useful. Since then, entanglement has been
viewed not as a curiosity, but as a real physical resource that can offer completely new
unexpected effects. The breakthrough was the discovery of dual effects, i.e. dense coding
and quantum teleportation in which the ebit plays a central role, i.e. a pair of
qubits in a maximally entangled state, distributed between the sender and receiver.
Remarkably both entanglement-based effects circumvent the non-cloning and Holevo
theorem.

\subsection{Super dense-coding}

Suppose Bob wants to send to Alice two bits of
information, using only one noiseless qubit.  According to Holevo's theorem, only one bit can be
transferred with one qubit. So Bob would need two qubits for this. Bennett and Wiesner showed \cite{BennettW}
that if Alice and Bob have one ebits then it is enough to send only one qubit to transmit one of the four messages
($00$,$01$,$10$,$11$) to
Alice. To do this, Bob encodes messages using local different unitary operations
$U_{00}$, $U_{01}$, $U_{10}$, $U_{11}$ on his qubit, generating orthogonal Bell
states (Bell base), and sends the qubit to Alice, which measures the combined two qubits. The four orthogonal Bell
states represent the four distinguishable messages. The first implementation of a super-dense photon encoding protocol was
made by Mattle et al. \cite{DensCodExpQC} in which Bob performed unitary operations using a combination of
half and quarter revolutions of the wavelet. The dense coding protocol was later implemented
in particular on atoms \cite{AtomDensCod} and nuclear magnetic resonance \cite{NMRDenCod}.

\subsection{Quantum teleportation}

The most astonishing prediction of quantum inscription is quantum teleportation -- a dual
effect to dense coding that demonstrates the remarkable power ``exotic'' combination quantum and classical resources
(see the fascinating story of the discovery \cite{QC_celebrTe}).

This time Alice wants to send one qubit to Bob in an unknown state, but not by physical qubit
transfer, having two classic bits at her disposal. Obviously, quantum information cannot be
transferred with classical bits.  Let now consider the
situation if we provide partners with 1 ebit of entanglement. Now Alice can perform a
measurement on her two particles, i.e. a qubit in an unknown state $\phi$ and a particle from the
entangled pair. It is not hard to see that this measurement is identical to what Bob made in
high-density coding. Alice gets one of four possible outcomes with a $\frac14$ probability:
00,01,10,11.  Having two
bits at her disposal, Alice can send information via the classical channel to Bob which of the
results she received. Depending on the result, Bob uses one of the transformations:
$U_{00}\cong I$, $U_{01} \cong \sigma_x$, $U_{10} \cong \sigma_y$, $U_{11} \cong \sigma_z$ where
$\sigma_x$, $\sigma_y$, $\sigma_z$, are standard Pauli operators. At this point, his
particle from the entangled pair it will be in state $\phi$. Note that Alice's measurement provides
no state information (the bits are completely random), but is part of a quantum operation. So
the transmission of the qubit had to take place immediately at the moment of Alice's
measurement. There is no conflict with special relativity here because quantum inscription
predicts that any operation on one subsystem does not cause measurable changes on the other
subsystem regardless of the state of the entire system. Note that there is no contradiction here
with the prohibition on cloning, since the initial state of the qubit was completely erased in
Alice's laboratory and then recreated, but not known in Bob's laboratory.
It should be finally stressed that here no information about the
unknown state $\phi$ is transferred via a classical channel that only
conveys the message about the recovery operation at Bob’s lab which is
completely independent on $\phi$.

Original teleportation protocol
was extended including to continuum variables \cite{Telep_con.quant.variab1, Telep_con.quant.variab2}. Quantum
teleportation, was demonstrated in pioneering experiments by the
Zeilinger \cite{Wieden-teleportation} and De Martini
\cite{Rzym-teleportation} teams.  Furusawa and co-workers
\cite{Kimble-teleportation} independently carried out a unconditional
teleportation on continuous variables (see in this context
\cite{Telep_con.quant.variab1, Telep_con.quant.variab2,Tel_telecom,
  Tel_Danube, tel100}). Later, quantum teleportation was
demonstrated in many beautiful experiments
\cite{QC_telep,Telepor_Teneryfe, Telep_single-atomQM,
  Telep_solid_state, Telepor_qutrit, Telepor_three_dimens}.  In 2017, a photon was teleported from
Ngari ground station to the Micius satellite (with an orbit from 500
to 1400) \cite{Micius,Tel_sat}.

Quantum teleportation has been continuously researched for more than 20 years (see ref. \cite{Entropy}) due to its
central role in the development of quantum information processing including quantum computing \cite{QC_telep,Salih},
the quantum internet and its relationship to the foundations of physics.
Various generalisations of the original protocol have been proposed.
In particular, the original protocol was generalized including
general teleportation channel \cite{H99}, multiport
teleportation \cite{Ishizaka2008, Ishizaka2009, Mozrzymas},
teleportation with multiple sender-receiver pairs \cite{Roy2020},
telecloning \cite{Murao1999}.

\subsection{Entanglement swapping}

The peculiarity of multi-particle entanglement is that one can entangle particles that
have never interacted with each other in the past. That such an effect may take place was suggested by the first Yurke
and Stoler (1992b) \cite{Yurke_Stoler}.  This idea was implemented in the pioneering paper: \emph{``Event-ready-detectors'' Bell
experiment via entanglement swapping}. In this scenario, arbitrarily distant partners Alice and Cecilia and Bob and
David share entangled EPR pairs of photons coming from independent sources:
\begin{eqnarray}
  |\Phi^+\rangle_{AC} = \frac1{\sqrt2} (|00\rangle +|11\rangle), \nonumber \\
  |\Phi^+\rangle_{BD} = \frac1{\sqrt2} (|00\rangle + |11\rangle)
\end{eqnarray}
The system is then described as
\begin{equation}
  |\Phi^+\rangle_{AC} \otimes |\Phi^+\rangle_{BD}
\end{equation}

Now Cecilia and Bob make a combined measurement in Bell's basis on $B$ and $C$ particles. As a
result, A and D particles become entangled even though they never interacted with each other. Note that this is
equivalent to teleporting entanglement of one EPR pair through the other. Soon the entanglement swapping was generalised to
multiparticle systems \cite{Bose}.  It provided the operational foundations of multi-photon interferometry, in particular
the method of interference of photon pairs from independent sources
(see review \cite{RMP_Zuk}).  The entanglement swapping \cite{EsEx,
  ESefficient, Swap2019} has found applications among other in the
generation of multi-photon entangled states \cite{ES_Multiphoton},
device-independent key distribution \cite{ESKD} and construction of
quantum repeaters \cite{Repeaters, Repeaters2001, Behera2019}, quantum photonic
\cite{EsPhotonic}, secret sharing \cite{ESsec, AdvESsec}.

\section{Detection of quantum entanglement}

All of the above effects and many other non-classical tasks based on quantum information
processing require high purity quantum entanglement. Unfortunately, this subtle resource is extremely sensitive to
interaction with the environment and it degrades very quickly, i.e. pure states change into mixed (noisy) states with
less entanglement.
This opened up important issues: how to theoretically check whether a
given state is entanglement and is it possible to detect noisy entanglement in the
laboratory?

In general, characterizing entangled states regardless of the measure of utility for
specific tasks is so-called NP difficult problem \cite{Gurvits2003}. The partial characterization was achieved using criteria
that provide the necessary but not sufficient conditions for deciding whether a state is entangled or not. The
breakthrough was the paper of Peres \cite{Peres1996}, who proposed an extremely strong separability test based on the partial
transposition operation. From mathematical point of view it is positive but not completely positive map thus
non-physical one. Such an operation is performed on one $S_1$ or $S_2$ of subsystem on complex state of the system
$S$. If the state subjected to such non-physical surgery does not survive in the sense that it will cease to be
positive and lose its probabilistic interpretation, then the state was entangled. Mathematically speaking, this means
that its partially transposed density matrix has at least one negative eigenvalue. Based on
the complete classification of positive mappings for low dimensions \cite{Woronowicz1976} it was proved
that the PPT condition is a necessary and sufficient condition for the separability of $2 \times 2$ and
$2 \times 3$ systems \cite{HHH96} which gives a complete characterization for low-dimensional states of systems. In general
necessary and sufficient, albeit non-operational, separability condition based on positive mappings was provided
\cite{RMP2009}.

\begin{figure*}
  \centering
  \includegraphics[width=6cm]{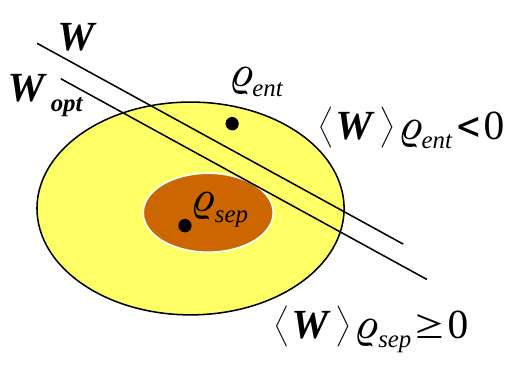}
  \caption{The line represents hyperplane corresponding to the
    entanglement witness $W$. All states located to the left of the
    hyperplane or belonging to it (in particular all separable states)
    provide nonnegative mean value of the witness, i.e.
    $\Tr(W \varrho_{sep}) \ge 0$ while those located to the right are entangled
    states detected by the witness. $W_{opt}$ is optimized
    entanglement witness.}
  \label{fig:3}
\end{figure*}

The above structural criteria based on positive non-physical mappings of the quantum state,
while strong, that they cannot be implemented in a laboratory. Fortunately, based on the geometric properties of convex
sets, it was possible to formulate a linear separability criterion that could be implemented physically.
Namely, from the convex set theory and the Hahn-Banach theorem, it follows that for any
entangled state $\varrho_{ent}$ there exists a hyperplane in the space of operators separating $\varrho_{ent}$
from the set of separable states $S$. Such a
hyperplane is defined uniquely by the Hermitian operator $W$ (entanglement witness) \cite{Terhal2000}. Then the state is entangled
iff expectation value $W$ on $\varrho_{ent}$ is negative i.e. $\langle W \rangle \varrho_{ent} < 0$ whereas its expectation
value on all separable states $\langle W \rangle \varrho _{sep} \geq 0$ (see Fig. \ref{fig:3}).  It was shown, that such a witness
can be optimized by shifting the hyperplane parallel to the set $S$ \cite{Lewenstein2000,Lewenstein2001}.
Thus the detection
of entanglement consists in measuring the mean value of a properly selected observable.
Remarkably there is a ``footbridge'' Jamiolkowski isomorphism \cite{Jamiolkowski}
which allow to go from nonphysical positive maps to the physical measurable quantities to
Hermitian operators (entanglement witness), which provides a necessary and sufficient
condition separability \cite{HHH96}.

The entanglement witness criterion has a number of advantages: i) it is universal in the sense
that for any entangled state always exist entanglement witness; ii) It certifies entanglement in
experiments in the presence of noise; iii) It allows to detect the presence of
entanglement even in several measurements in contrast to tomography, where the number of
measurements increases exponentially with the number of particles. The disadvantage
is that the witness must be precisely selected for the examined state.  The
quantum entanglement detection based on entanglement witnesses has found wide
applications for the certification of two- and multi-partite states
\cite{Barbieri2003, Bourennane2004, Roos2004, Altepeter2005,
  Hffner2005, Mikami2005, Langford2005, Laskowski2012, Dirkse2020}
in different physical scenarios.
Interestingly, the concept of measurement-device-independent entanglement witness which allow
one to demonstrate entanglement of all entangled quantum states with untrusted
measurement apparatuses was introduced \cite{Gisin_MDIEW}.

The theory of entanglement detection was developed in different directions \cite{RMP2009,Kus_entDetec13}. The
other separability criteria based on correlation tensor was proposed \cite{Badziag, Laskowski, Markiewicz}
for bipartite and multipartite
scenario.  Recently it has been proved that or enhanced nonlinear realignment criterion \cite{Zhang} is equivalent to the
family of linear separability criteria based on correlation tensor i.e. the family of (linear) entanglement witnesses
\cite{Sarbicki}. It was also demonstrated that the separability criteria based of the correlation tensor are weaker than positive
partial transposition criterion \cite{Chruscinski2020}.

\section{Entanglement distillation and bound entanglement}

After Peres discovered the entanglement criterion of partial transposition, a problem
arose. If the state was untangled, it will remain untangled after the partial transposition operation. It was natural
to ask are there states in nature that are entangled and have positive partial transposition? When such states were
discovered in 1997 \cite{Pawel_separ97}  they were treated as a mathematical singularity with no reference to physical reality.
At about the same time, Bennett and colleagues were working on the problem of how to reverse the entanglement
degradation process. In 1996, they published a paper that played a key role in the
theory of entanglement manipulation
\cite{BBPSSW_pur} (see also \cite{PrivacyAmpl}).  Namely, they introduced a natural class of entanglement manipulation operations by
experimentalists in distant laboratories: the two partners can perform any local operations on their entangled
particles and communicate via the classical channel (LOCC).
Consequently, they introduced the entanglement distillation protocol:
The partners share n copies of the $\varrho_{AB}$
state which contains noisy entanglement. With the help of local quantum operations and classical communication, they
determine a smaller number of $m$ ($m < n$) of almost maximally entangled pairs -- two-qubit singlet states
$|\Psi^-\rangle_{AB}$.
When the protocol is optimal, the constant $m / n = D$ is a measure of entanglement in a noisy state $\varrho$ (distillable
entanglement).

Distillation protocol raised the natural question: Can all noisy states be
distilled in this way? It turned out that all noisy entangled two-qubit states can be distilled \cite{Horodecki_distill}.
It was a big surprise that the distillation protocol does not
work for the higher dimension systems \cite{Bound_ent, Terhal_PToday}. It turned out that
the environment can contaminate pure entanglement in such a
way that it is no longer possible to recover it by distillation with LOCC.
Thus, the  entangled states with positive partial transposition are non-distillable.
Thus, in Nature there are at least two types of noise entanglement: free, that is, distillable
entanglement, and bound entanglement that cannot be distilled with LOCC \cite{Europhys_News}.
After 12 years, several centers simultaneously created the bound entanglement in the
laboratory on the photons \cite{Bournnane_bound, Piani_bound}, on ions
\cite{Blatt_bound}, in liquid in NMR \cite{Kampermann_bound}, with
light in continuous variable \cite{DiGuglielmo_bound} regime.

It has been shown that the bound entanglement is not a rare phenomenon , since its
presence was detected in thermal spin systems \cite{Toth_bound, Ferraro_bound}. Another surprise was that the
bound entanglement can be activated \cite{Bound_activ} and that a cryptographic
key can be extracted from bound entangled states it \cite{Key_bound}.
The latter lead to the general paradigm for distilling classical key from quantum
states in terms of so called private bits (P-bits) \cite{P-bits} [See experimental
implementation  \cite{Key_bound_exp})].  Moreover bound entangled states can violate Bell inequalities \cite{PeresConject}
and can be useful in quantum metrology
\cite{QMetrH_limin,BoundStatMetr, BoundQMetr}. Another an interesting open problem is the
use of bound entanglement states in the device-independent quantum key
distribution \cite{ArnonFriedman2018, ArnonRateQKD, ChrisRateQKD}.

\section{Breaking the classical order}

When analysing the structure of entangled states, Schrödinger noticed another peculiarity of
quantum correlations that astonished him, as evidenced by the three question marks that appear in his unpublished notes
in 1932 [Note in arxiv].  In 1935, he makes a laconic conclusion: ``Best possible knowledge of a whole does not include
best possible knowledge of its parts -- and that is what keeps coming back to haunt us.'' \cite{Schrodinger}. It was very
disturbing because it meant breaking the classical order in complex systems. As is known in the classical world, the
measure of the randomness (disorder) of an individual random variable X is the Shannon entropy:
\begin{equation}
  H(X)= - \sum_i p_i \log p_i
\end{equation}
where $p_i$ -- probabilities of events, $\sum p_i = 1$.

For two random variables $X$ and $Y$, the total Shannon entropy is $H (X, Y) = \sum_{ij} p_{ij} \log p_{ij}$
and conditional entropies $H(X|Y)$, $H(Y|X)$ are always:
\begin{equation}
  H(X|Y) \equiv H(X,Y) - H(Y) \ge 0, \quad H(Y|X) \ge 0
  \label{eq:H-ineq}
\end{equation}
which shows that the entropy of a subsystems $H(X)$, $H(Y)$ never exceeds the total entropy of
the system $H(X,Y)$.

In the quantum world, the measure of quantum disorder is the von Neumann entropy $S(\varrho)$ 
defined for the state $\varrho$:
\begin{equation}
S(\varrho) = - \Tr(\varrho \log \varrho) = \sum \langle \phi_i | \varrho \log \varrho | \phi_{i} \rangle
\end{equation}
where $\{ \phi_i \}$ any complete orthogonal system in $H$.
When density matrix $\varrho$ is diagonal, it
can be regarded as a quantum counterpart of a classical discrete probability distribution
as a natural description of quantum information source. Then von Neumann entropy can
be written in a form similar to the Shannon entropy
\begin{equation}
  S(\varrho) = - \sum_i p_i \log p_i,  
\end{equation}
where the quantum probabilities $p_i$ are the eigenvalues of the operator $\varrho$
satisfy $\sum p_i = 1$.

The Schrödinger observation was quantified using the von Neumann
entropy \cite{H94, Cerf97}. It has been proved that the entropy of the
subsystem $A$ or $B$ can be greater than the entropy of the entire
system $AB$ only when the system is in a entangled state.  This
implies that quantum conditional entropies
$S(A|B) \equiv S(AB) - S(B)$, $S(B|A)$ can be negative, which means
that the disorder in the whole AB system may be smaller than in the
subsystems A or B. Recalling our example with photons entangled in
polarization, we can see that everything happens agrees. The
polarizations of the photons measured in the laboratories of Alice and
Bob are completely random, while the entangled pair is in perfect
order.  Thus entanglement can break the classical order which is the
source of the informational ``paradox'' of Schrödinger.

\section{Negative information in quantum communication}

The breaking of the classical order was both intriguing and incomprehensible, especially
in the context of Shannon's theory, in view of the fact that the negativity of quantum
conditional entropy had no operational significance.  Let us recall
that at the heart of the classical Shannon communication theory is the theorem of noiseless
coding, which says that a necessary and sufficient number of bits for faithful transmission is equal to Shannon's
entropy $H$  \cite{Shannon}. Schumacher showed that if in Shannon's theory we replace messages by quantum states and bits by qubits,
then the necessary and sufficient number of qubits for faithful transmission is equal to the von Neumann entropy $S(\varrho)$ \cite{S_qubit}.
Soon after Schumacher and Westmoreland \cite{Westmoreland97}
and Holevo \cite{Holevo98} generalized Shannon’s channel coding
theorem.
Three kinds of quantum channel capacities was introduced: classical,
quantum and private capacity, which play an important role in quantum
communication \cite{Bennett_capacity, Devetak_cap_IEEE,
  Bennett_Dev_capacity, Horodecki_private-cap-IEEE,
  Winter_private-cap}.
The essential difference between the last two capacities \\
is the following: The quantum capacity is achieved in the process which guarantees that \\
information in any basis stays uncorrelated from the environment after the transfer (which may \\
be shown to be equivalent to BB84 paradigm). Remarkably in the definition private capacity \\
much more relaxed condition is required: only  one base is needed to stay uncorrelated in the \\
above sense. Note that the private capacity while in general higher than the \\
quantum one may have subject to severe restriction in quantum repeater scenario \cite{Buml2015}
(see more \cite{Wilde}).

Meanwhile, for a long time there was no quantum counterpart of Slepian-Wolf theorem \cite{Slepian}.  Namely in
1973 Slepian and Wolf formulated in framework of classical communication the following problem: The two partners
Alice and Bob have random variables $X$ and $Y$ that are correlated with each other. Bob is given some incomplete
information of $Y$ in advance. Alice is in possession of the missing information of $X$. Bob's job is to obtain the
missing information of $X$. The question is how much additional information Alice has to send to her partner. Slepian and
Wolf proved that the amount of information that Bob needs is expressed by the conditional entropy: $H (X | Y) \equiv H (XY) - H (Y)$
which is a measure of the partial information that Alice must send to Bob. This quantity is
always positive.

In 2005, Horodecki et al. \cite{Horodecki2005} proposed a quantum version of the above scenario: Alice and
Bob have a system in some unknown quantum state $\varrho_{AB}$ which contains the complete information.
Bob has some information about state $\varrho_B$, while Alice has the missing information $\varrho_A$.
The task is as follows: how much information does Alice have to send to Bob for him to have complete information.
The quantum equivalent of the Slepian Wolf theorem says that this quantity is given by the von Neumann quantum
conditional entropy:
\begin{equation}
  \label{eq:18}
  S(A | B) \equiv S(AB) - S(B)
\end{equation}
where $S(B)$ is the entropy of the Bob state while $S(AB)$ is the entropy of the cumulative
$\varrho_{AB}$ state. Contrary to the classical
conditional entropy $H(X | Y)$, the conditional entropy can be both positive and
negative. Conditional quantum entropy has an operational interpretation of missing information:
If $S(A | B)$ is positive -- this is the missing information that Alice must send to
Bob via qubits (classical analogue). If $S (A | B)$ negative, Alice does not need to send the missing
information via qubits. Additionally, Bob and Alice get free ``quantum impulses'' to send
a certain number of qubits in the future, for example for teleportation.

Finally it should be stressed that the above analysis is a strong completion of the previous result \cite{DevetakWinter}
which says that for any state with the quantity \eqref{eq:18} negative there exists an entanglement distillation protocol
with one way classical communication (from Alice to Bob) that achieves the number of e-bits per input noisy
pair given by \eqref{eq:18}.

\section{Entropy inequalities -- nonlinear witnesses of entanglement}

Von Neumann entropy can be generalised to the R{\'e}nyi family $\alpha$-entropy
$S_\alpha(\varrho)$
\begin{equation}
  S_\alpha(\varrho) = \frac1{1 - \alpha} \ln \Tr \varrho^\alpha, \quad \alpha > 1
\end{equation}

It is easy to check that the R{\'e}nyi entropy in the $\alpha \to 1$ limit turns into
the von Neumann entropy $S(\varrho)$.
The natural question was whether there are quantum states that satisfy the analog of classical
inequalities \eqref{eq:H-ineq}. In 1996 \cite{RMH_96} it was proved
that all non-entangled (separable) states at a finite dimensional
Hilbert space for $\alpha = 1.2$ satisfy $\alpha$-entropic inequalities:
\begin{eqnarray}
  S_\alpha (A | B) = S_\alpha (\varrho_{AB}) - S_\alpha (\varrho_{B}) \geq 0,  \nonumber \\
  S_\alpha (B | A) = S_\alpha (\varrho_{AB}) - S_\alpha (\varrho_{A}) \geq 0
\end{eqnarray}

It presents entropic nonlinear entanglement criterion which does not require a priori
knowledge of the state.

Nonlinear experimentally friendly collective entanglement witnesses were also proposed, which
also do not require prior knowledge of a given state \cite{PH_Ekert, PH2003}.
In \cite{NonlinExp} Bovino et al.
demonstrated first experimental measurement of a non-linear entanglement witness
$S_2 (\varrho) = - \Tr \ln \varrho^2$, using local measurement on two pairs of polarization entangled
photons.

At first, it seemed that the entropy criterion based on nonlinear entanglement witnesses,
generally weaker than the criterion based on linear ones, will not play a major role.
However, it turned out
that, the feature of non-linearity is its strength. In particular, the nonlinear entanglement witnesses ``feel'' the
subtle features of entanglement in quantum multi-body systems. In last decade there has been a renaissance of entropic
witnesses opening up the field for wide applications. For pure or nearly pure states, entanglement was detected
using R{\'e}nyi $S_2$ entropy via a multi-body quantum
interference \cite{Islam_entr_multi_body,Kauf_entr_multi_body,Linke_entr_multi_body, Alves_entr_multi_body, Daley_entr_multi_body}
and local random measurements \cite{Brydges_loc_rand, Enk_loc_rand, Elben_loc_rand, Elben_loc_rand_2019,
  Huang_loc_rand}. An experimental measurement of nonlinear witnesses of
collective entanglement using hyper-entangled two-quart states has
been performed \cite{Travnicek_esp_non_wit}, see also
\cite{Bartkiewicz_esp_non_wit}.
Quite recently, an experimental multi-body mixed state detection method has been proposed based on the positive
partial transposition of a density matrix condition. This protocol gives the first direct PT measurement of moments in
a multi-body system \cite{Preskill_2020}.

\section{Quantum parallelism as the basis for quantum computing}

Quantum computing is processing information using sequence of unitary
operations (quantum gates) in order to obtain an answer to a
predetermined question, e.g. is a given number factorizable with high
probability \cite{Barenco1995}.  As we have seen
single qubit allows two basic states to be stored and processed
simultaneously. The problem is that the decoherence process being a
result of disturbance by environment occurs within
a short time (decoherence time) destroys coherence. Roughly speaking
decoherence time is the characteristic time for a generic qubit state (2) to be
transformed into the mixture $\varrho = |a|^2 |0\rangle\langle0| + |b|^2 |1\rangle\langle1|$. One of the basic conditions
for effective quantum computing requires that long relevant decoherence times, much
longer than the gate operation time. This is one of the five basic DiVincenzo criteria
required for a physical implementation of quantum computing \cite{DiVincenzo_criteria}. If we
take a superposition of n qubits then a pure state will represent a simultaneous
superposition of $N=2^n$ possible distinct basic states.
\begin{equation}
  |\Psi\rangle = \sum_{i=0}^{N-1} C_i |i\rangle
\end{equation}

It is remarkable, that one can processes simultaneously an exponential number of basic
states. This feature (quantum
parallelism) underlies the superiority of quantum computing over classical one.
To illustrate the latter suppose that we have access to quantum oracle that computes a given function
$f(i)$ from an input $i$ of $n$ qubits ($i=0,1 \ldots 2^n$).

Having a prepared string of qubits in the fiducial state of 0 and applying to each qubit,
in parallel, Hadamard gate, we obtain a register of n qubits in an equal superposition of all bit strings
\begin{equation}
  H |0\rangle \otimes H |0\rangle \otimes \cdots \otimes  H |0\rangle
  = \frac1{\sqrt{N}} \sum_{i=0}^{N-1} |i\rangle
\end{equation}
where $|i\rangle$ is the computational basis state indexed by the binary number
that would correspond to the number $i$ in base-10 notation.

Now suppose that the function $f$ is evaluated by unitary transformation
$U_f: |x\rangle|0\rangle \to |x\rangle|f(x)\rangle$. Then the linearity of quantum formalism implies
\begin{equation}
  U_f: \frac1{\sqrt{N}} \sum_{i=0}^{N-1} |i\rangle |0\rangle
  \quad \to \quad \frac1{\sqrt{N}} \sum_{i=0}^{N-1} |i\rangle |f(0)\rangle
\end{equation}

This means that all possible evaluations of the function $f(x)$ can be
evaluated in a single step.

The idea of quantum computing received a lot of support when it was
discovered that certain difficult computational problems such as
number factoring (Shor's algorithm \cite{Shor_alghor}) and searching unstructured data
(Grover's algorithm \cite{Grover_97}) can be solved far more efficiently than classically.
The efficiency of computation is measured by the computation
complexity that is number of steps required to solve a given task as a function of the size of the input. The
important algorithms: Deutsch-Jozsa \cite{Deutsch_Jozsa}, Shor \cite{Shor_alghor} and Grover \cite{Grover_97} have been discovered that
demonstrate quantum supremacy over classical computing. All three algorithms have be implemented on primitive quantum
computers based on NMR techniques \cite{IBM_Computer}, in ions traps \cite{ExpAlgh_G_ion} and quantum dots \cite{Exp_DJ_dots}.
Since then many other algorithms have been discovered, such as quantum simulations, \cite{AlgQS}
variational quantum solvers \cite{AlgWQS} which demonstrate quantum supremacy [see more \cite{QCS}].

Any realistic implementation of universal quantum computation is big challenge. It must meet
the DiVincenzo criteria \cite{DiVincenzo_criteria}.  Except decoherence criterion,
there are four more:
\begin{enumerate}
\item A scalable physical system with well characterized qubits.
\item The ability to initialize the state of the qubits to a simple
  fiducial state.
\item A ``universal'' set of quantum gates.
\item A qubit-specific measurement capability.
\end{enumerate}
Notoriously the quantum
computing process is disturbed by the interaction with the environment, causing the occurrence of errors. Therefore
both bit (0,1) and phase (``$0+1$'', ``$0-1$'')  they must be protected. This seems impossible due
to the non-cloning theorem. Fortunately, Shor \cite{Shor} and Steane \cite{SteanCodes} overcame this difficulty by introduction
of the error correction codes. The trick is that the information of a one logical qubit can be spread onto a highly entangled
state of several physical qubits.
\begin{eqnarray}
  &&|0\rangle \to |\boldsymbol{0}\rangle_{\boldsymbol{L}}
     = [(|000\rangle +|111\rangle) (|000\rangle +|111\rangle) \nonumber \\
  &&\hspace{0.5cm}\times (|000\rangle +|111\rangle)]/2\sqrt2 \\
  &&|1\rangle \to |\boldsymbol{1}\rangle_{\boldsymbol{L}}
     = [(|000\rangle -|111\rangle) (|000\rangle -|111\rangle) \nonumber \\
  &&\hspace{0.5cm}\times (|000\rangle -|111\rangle))]/2\sqrt2
\end{eqnarray}

This code first introduced by Shor \cite{Shor} corrects both bit error $\sigma_x$ and phase error $\sigma_z$.

Of course the error correction procedure itself is not error-free. Fortunately the
possibility of efficient quantum computing is based on the so-called the threshold
theorem:  Error probability of elementary
operation smaller than some threshold value $p < p'$ then efficient quantum computing possible
\cite{Aharonov_Ben-Or_FT, Knill_resilient_QC}. In practice, this condition, which is the basis of efficient quantum
computing, is extremely demanding.
Already in 1995 it was demonstrated that the quantum computing can be
implemented with cold ions confined in linear trap and interacting
laser beams \cite{Cirac1995}.
The first 7-qubit quantum computer from IBM and Stanford University based on
nuclear magnetic resonance realized Shor's algorithm, decomposition of the number 15 = 3x5 \cite{IBM_Computer}. The scale
of the difficulties is evidenced by the fact that a qualitative breakthrough in this field took place only after 18
years. Namely researchers at Google’s quantum-computing laboratory in Santa Barbara, California, announced the
first-ever demonstration of quantum supremacy on the 53 qubit quantum computer Sycomore, made of superconducting
circuits that are kept at ultracold temperatures \cite{Sycamore_Comput}. It executes algorithms quantum with 1500 gates. It
is impressive achievement, however, it was designed for a specific problem –- boson sampling \cite{Boson_sampling},
which is simplified non-universal model for quantum computing that may hold the
key to implementing the first ever post-classical quantum computer.
More specifically this is the process in which a very nontrivial
output statistic is achieved quantumly which requires (under some
reasonable assumptions) exponentially longer time to be produced by
classical machines. While it is not a quantum algorithm in a standard
form its remarkable practical application to fast finding of some
graph properties are predicted.

Quite recently Jian-Wei Pan and colleagues at the University of Science and Technology of China
in Hefei et al. announced in December 2020 photon-based quantum computer, which demonstrates quantum supremacy via
boson-sampling with 50-70 detected photons \cite{computer-China}.  It could find solutions to the boson-sampling problem
in 200 seconds, while classical China’s Taihu-Light supercomputer. need 2.5 billion years. In in contrast to
Google’s Sycamore, the Chinese team’s photonic circuit is not programmable \cite{Ball_comp.China}.

\begin{figure*}
  \centering
  \includegraphics[width=11cm]{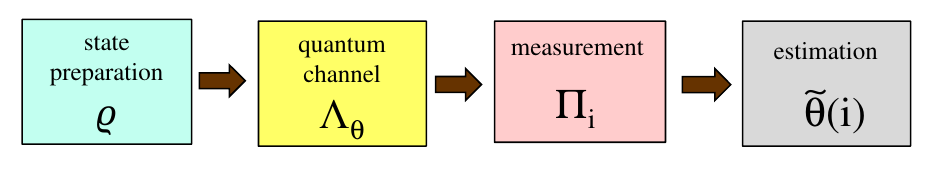}
  \caption{Phase estimation scheme}
  \label{fig:4}
\end{figure*}

\section{Entanglement -- resource in quantum metrology}

The discovery that the use of entangled states in quantum metrology can improve the
precision of measurements \cite{Caves_81, Giov.Lloyd} led to the rapid development of quantum enhanced
metrology \cite{RMP_metr} which allows measure physical quantities by estimating the phase shift $\theta$
using interferometric techniques. A basic problem in quantum metrology can be
formulated as in the diagram (Fig. \ref{fig:4}): A probe state $\varrho$ of $N$ particles is prepared and then
subject to a parameter-dependent quantum channel $\Lambda_\theta$. The state $\varrho_\theta = \Lambda_\theta (\varrho)$ is finally
measured via POVM measurement $\{ \Pi_i \}_I$. It provides conditional probability distribution
$p(i|\theta)$, which is used to estimate of $\theta$ via estimator function $\tilde\Theta(i)$ for given the
measurement outcome $i$. The estimation of the phase shift is limited by uncertainty:
\begin{equation}
  \label{eq:26}
  \Delta^2\tilde\theta = \langle (\tilde\theta - \theta)^2 \rangle
\end{equation}
where $\langle\;\rangle$ means the average over all measurement results. The task is to find the
optimal probe state $\varrho$, the optimal measurement $\Pi$ and estimator , which minimize the
uncertainty.  For unbiased estimators and m independent measurements the phase
uncertainty is limited by the quantum Cramer-Rao bound:
\begin{equation}
  \Delta\tilde\theta \ge \frac{1}{\sqrt{mF_Q(\varrho_\theta)}}
\end{equation}
where $F_Q$ is quantum Fisher information which quantifies asymptotic usefulness of
quantum state and it can be estimated for the different quantum channels \cite{RMP_metr}.

For unitary and noiseless quantum channel
$\varrho_\theta = \Lambda_\theta (\varrho) = e^{-iH \theta} \varrho \, e^{+iH \theta}$ the quantum Fisher
information optimized over measurement can be expressed in the form:
\begin{equation}
  F_Q[\varrho,H] = 2 \sum_{k,l}\frac{(\lambda_k-\lambda_l)^2}{\lambda_k + \lambda_l} |\langle k| H | l\rangle|^2
  \label{eq:fisher-inf}
\end{equation}
where $H$ is the generator of the phase shift of the system, and
$\varrho = \sum_k \lambda_k |k\rangle\langle k|$,
$\sum_k\lambda_k=1$.

For unitary dynamics of the linear two-mode interferometer the generator of the phase
shift is $H \equiv \boldsymbol{J}_{\vec{n}} = \vec{n} \cdot \boldsymbol{J}$ where $\boldsymbol{J}_{\vec{n}}$
is a component of the collective spin operator angular
momentum in the direction $\vec{n}$.  It has been shown
\cite{Lloyd_2006,Pezze_2009}, that for the separable input
{$N$-particle} states, the quantum Fisher information is bounded by
$F_Q [ \varrho_{sep}, \boldsymbol{J}_{\vec{n}} ] \le N$. Hence the
phase uncertainty $\Delta\tilde\theta$ is bounded by standard quantum limit (SQL)
$\Delta \theta_{SN} : \Delta \tilde\theta \ge \Delta \theta_{SN}$ where
\begin{equation}
 \Delta \theta_{SN} = \frac{1}{\sqrt{mN}}
\end{equation}
By using entangled probe states it is possible to overcome the SQL \cite{RMP_metr}. Quantum
formalism imposes fundamental constraints on measurement precision that scales like $1/N$.
It has been shown that, for general probe states of N particles $F_Q$ is bounded by
$F_Q [ \varrho, \boldsymbol{J}_{\vec{n}} ] \le N^2$, \cite{Lloyd_2006,Pezze_2009}
and this inequality can be saturated by certain maximally entangled
states. It allows to obtain optimal Heisenberg bound for the phase uncertainty
\begin{equation}
  \Delta\theta_{HN} = \frac{1}{\sqrt{m}N}
\end{equation}
Note that the genuine multipartite entanglement is needed for reaching the highest
sensitivities in some metrological tasks using two-mode linear interferometer \cite{Toth,
Laskowski_metr, Toth_Apellaniz}.
Recently, various experiments have demonstrated beating the
SQL (see \cite{Xie} and references there in).

In a realistic scenario, quantum phase estimation requires taking into account the
influence effects of losses and decoherence \cite{Huelga_97, Demkowicz_2009,
Matsuzaki_2011, Kolodynski, Chin_2012, Demkowicz_2017, Preskill_2018,
Chabuda_2020}.  In particular for $N$ probe particles prepared
in state $\varrho^N$ and noisy channel $\Lambda_\Theta^{\otimes N}$ , that acts independently on each particle
$\varrho_\theta^N = \Lambda_\Theta^{\otimes N} (\varrho^N)$, quantum Fisher information $F_Q (\varrho_\theta^N)$
has asymptotically in $N$ a bound that scales linearly with $N$: $F_Q (\varrho_\theta^N) \le N\alpha$ giving bound \cite{Kolodynski}:
\begin{equation}
  \Delta\tilde\theta \ge \frac{1}{\sqrt{\alpha m N}},
\end{equation}
where $\alpha$ is constant. Thus the
supremacy over SQL is only limited to constants factor. In particular, in the optical
interferometry with losses for a generic two mode input $N$-photon state with precisely
defined total photon number $N$ the limit of phase sensitivity is:
\begin{equation}
  \Delta\tilde\theta \ge \sqrt{\frac{1 - \eta}{\eta N}}
\end{equation}
where $\eta$ is optical transfer coefficient. This bound generalized to states having
uncertainty photon number such as coherent states and squeezed states was used to
estimate the fundamental bound on GEO 600 interferometer strain sensitivity \cite{GEO_600} where
the phase noise decoherence \cite{Franzen}, and quantum back-action
are negligible \cite{Caves_81}. It has been
shown that the coherent-state squeezed vacuum strategy is optimal one for phase
estimation with high precision on absolute scale \cite{GEO_600}.

Recently, a framework for optimization of quantum metrological protocols based on the
tensor network approach for the channel with the correlated noise and the phase
parameter unitarily encoded were presented \cite{Chabuda_2020}. Note that multiparameter estimation theory offers a general framework to
explore imaging techniques beyond the Rayleigh limit \cite{Parniak}.

Overall, the relationship between quantum metrology and the structure of quantum
states is still not entirely clear. For example there are very weakly entangled states
(bound entangled states) metrologically useful
\cite{QMetrH_limin,BoundStatMetr} as well highly entangled states that are
not useful for metrology \cite{Hyllus2010}.  It leads to the question: Are there situations were some
synergy effects occurs possibly with analogy to previous communication protocols such
as activation of bound entanglement? In attempt to answer this question, the criterion of
metrological usefulness have been proposed as follows \cite{Toth_H}:

The state $\varrho$ is metrologically useful iff there exists Hamiltonian $H$ such that Fisher
quantum information \eqref{eq:fisher-inf} is sharply greater than Fisher information for separable states
$F_Q [ \varrho_{sep} , H ]$ maximized over all separable states:
\begin{equation}
  F_Q [ \varrho , H ] > \max_{\varrho_{sep}}=F_Q[\varrho_{sep}, H] =: F_Q^{(sep)}(H)
\end{equation}
Then the metrological gain with respect to the Hamiltonian $H$ defines as
$g_H(\varrho) = F_Q[\varrho,H] / F_Q^{(sep)} (H)$ leads to the optimal gain
$g(\varrho) = \max_{local H} g_H(\varrho)$.
Having such defined
metrological usefulness it has been shown that the bipartite entangled states that cannot
outperform separable states in any linear interferometer, however can still be more
useful than separable ones if several copies of them are considered or an ancilla is added
to the quantum system. In particular it has been proved that all entangled bipartite pure
states are metrologically useful.

\section{Final remarks}

In this article, I have focused only on selected aspects of quantum
information. There are many other fascinating phenomena that deserve
presentation. These include quantum correlations beyond entanglement
\cite{Modi_RMP, Bennett_Horodecki},
nonlocality without
entanglement \cite{Nonlocality}, quantum channel super activation
effect \cite{Smith, Modi_RMP}, locking classical correlations in quantum states
\cite{Locking}, resources theoretical approach to
quantum thermodynamics \cite{Gour_RMP}, quantum Darwinism \cite{Zurek,
  Brandao, Le},  objectivity \cite{Korbicz, Horodecki, Pan,
  Scandolo_obj}, quantum based randomness amplification against
postquantum attacks \cite{ColbeckRenner, Gallego, Brandao_Random} and
others.
They all underline the extremely complex
nature of quantum information, which is not yet fully understood and
provokes many open questions (see for example \cite{five-problems}).
Among others there is a long-standing
question: If the quantum formalism can be consistently extend to
include quantum gravitation effect? If so, how it will impact on the
quantum information concept?

\section*{Acknowledgments}

I would like to thank J. Mostowski, and A. Wysmołek for encouraging me
to write this article based on the lecture given at the Extraordinary
Congress of Polish Physicists on the occasion of the centenary of the
Polish Physical Society. I would like also to thank J. Horodecka and
Ł. Pankowski for their help in editing of this paper. I
acknowledge support by the Foundation for Polish Science through IRAP
project cofinanced by the EU within the Smart Growth Operational
Programme (Contract No. 2018/MAB/5).

\medskip

\bibliographystyle{appa}
\bibliography{quantum-information.bib}

\end{multicols}

\end{document}